\documentclass[12pt]{iopart}

\usepackage{graphicx}

\usepackage[]{cite}

\usepackage[]{amssymb}
\usepackage[]{mathrsfs}
\usepackage[]{eucal}
\usepackage[]{tensor}
\usepackage[]{amsthm}
\usepackage[]{bm}

\newcommand{\dd}{\partial}
\newcommand{\df}{\mathrm{d}}
\newcommand{\w}{\wedge}
\newcommand{\Lie}{\pounds}
\newcommand{\nab}[1]{\nabla_{\!#1}}

\newcommand{\qqd}{\ , \quad}
\newcommand{\bc}{\begin{center}}
\newcommand{\ec}{\end{center}}
\newcommand{\be}{\begin{equation}}
\newcommand{\ee}{\end{equation}}
\newcommand{\bea}{\begin{eqnarray}}
\newcommand{\eea}{\end{eqnarray}}

\newcommand{\FF}{\mathcal{F}}
\newcommand{\GG}{\mathcal{G}}
\newcommand{\LL}{\mathscr{L}}

\usepackage[unicode]{hyperref}
\usepackage{xcolor}
\definecolor{pastgreen}{HTML}{669900}
\definecolor{pastblue}{HTML}{336699}
\definecolor{pastred}{HTML}{990000}
\definecolor{linkcol}{HTML}{663333}
\hypersetup{colorlinks,linkcolor={pastblue},citecolor={pastblue},urlcolor={pastblue}}

\theoremstyle{plain} \newtheorem{tm}{Theorem}[section]
\theoremstyle{plain} \newtheorem{lm}[tm]{Lemma}
\theoremstyle{definition} \newtheorem{defn}[tm]{Definition}
\theoremstyle{definition} 
\newcommand{\btm}{\begin{tm}}
\newcommand{\etm}{\end{tm}}
\newcommand{\blm}{\begin{lm}}
\newcommand{\elm}{\end{lm}}
\newcommand{\bdefn}{\begin{defn}}
\newcommand{\edefn}{\end{defn}}

\begin{document}

\begin{flushright}
ZTF-EP-19-03
\end{flushright}

\title[Schwarzschild spacetime immersed in test nonlinear electromagnetic fields]{Schwarzschild spacetime immersed in test nonlinear electromagnetic fields}

\author{Ana Bokuli\'c and Ivica Smoli\'c}
\address{Department of Physics, Faculty of Science, University of Zagreb, Bijeni\v cka cesta 32, 10000 Zagreb, Croatia}
\eads{\mailto{abokulic@phy.hr}, \mailto{ismolic@phy.hr}}

\date{\today}

\begin{abstract}
Kerr black hole immersed in test, asymptotically homogeneous magnetic field, aligned along the symmetry axis, is described by Wald's solution. We show how the static case of this solution may be generalized for nonlinear electromagnetic models via perturbative approach. Using this technique we find the lowest order correction to Wald's solution on Schwarzschild spacetime in Euler--Heisenberg and Born--Infeld theories. Finally, we discuss the problem of highly conducting star in asymptotically homogeneous magnetic field.
\end{abstract}


\vspace{2pc}


\section{Introduction}

Astrophysical black holes are surrounded by electromagnetic fields, produced by accompanying accretion disk or a wider galactic environment. It is believed that some of these fields play a crucial role in formation of powerful jets, ejected from galactic centres by supermassive black holes. Apart from this, electromagnetic potentials and charges appear among the variables of black hole thermodynamics, representing a meeting point of gravity, thermodynamics and gauge theories. Therefore, the study of black hole electrodynamics spreads across the wide spectrum between phenomenological physics and academic, purely conceptual research.

\medskip

Papapetrou has noticed \cite{Papa66} back in the 1960s that Killing vector fields, taken as gauge fields, satisfy the source-free Maxwell's equations. Namely, if $K^a$ is a Killing vector field of a spacetime $(M,g_{ab})$, then the 2--form $F = \df K$ immediately satisfies $\df F = 0$. Furthermore, by the Killing lemma \cite{Wald} we have
\be
\nab{b}\nabla^b K^a = \tensor{R}{^a^b_b_c} K^c = -\tensor{R}{^a_c} K^c \ .
\ee
So, if the spacetime metric $g_{ab}$ is a solution of vacuum Einstein field equation $R_{ab} = 0$, it follows that $\df\,{*F} = 0$ as well. In other words, such 2--form $F_{ab}$ represents a test electromagnetic field, solution of the source-free Maxwell's equations on the spacetime $(M,g_{ab})$. 

\medskip

Using this observation, Wald \cite{Wald74} has found a solution representing a simplified analytical model of the natural black hole environment: Kerr black hole immersed in a magnetic field which is asymptotically homogeneous and aligned with the axis of symmetry of the black hole. Suppose that $k = \dd/\dd t$ is stationary Killing vector field, $m = \dd/\dd\varphi$ axial Killing vector field and a constant $B_\infty$ magnetic field strength at infinity. Then Wald's solution in a spacetime of Kerr black hole with mass $M$ and angular momentum $J$ is given by
\be
F = \frac{1}{2}\,B_\infty \left( 2a\,\df k + \df m \right)
\ee
where $a = J/M$. Normalization is chosen such that both corresponding Komar electric and magnetic charges \cite{Heusler} evaluated on sphere at infinity vanish,
\be
Q_\infty = \frac{1}{4\pi} \oint_{\mathcal{S}_\infty} {*F} = B_\infty (-2aM + 2J) = 0 \ ,
\ee
and
\be
P_\infty = \frac{1}{4\pi} \oint_{\mathcal{S}_\infty} F = 0 \ .
\ee

\medskip

In this paper we go one step further, by looking at nonlinear modifications of the classical Maxwell's electrodynamics. Two earliest models of nonlinear electrodynamics (NLE) appeared back in 1930s: a phenomenological one proposed by Max Born and Leopold Infeld \cite{Born34,BI34} and a 1-loop QED correction calculated by Werner K.~Heisenberg and Hans H.~Euler \cite{HE36}. Curiously enough, Born--Infeld theory reappeared half century later in low energy limits of the string theory \cite{FT85,SW99}. Paradigmatic particle process which reveals nonlinearities in electromagnetic interaction is the ``light--by--light'', $\gamma\gamma \to \gamma\gamma$, scattering. The first direct experimental evidence of this process has been recently found by the ATLAS Collaboration \cite{ATLAS17}, via measurement of colliding ultra-relativistic lead ions at the Large Hadron Collider (an overview of earlier experimental constraints on NLE models can be found in \cite{FBR16}). Further analyses \cite{EMY17,AM18} of these results have strengthened the constraints on dimensionful parameter of Born--Infeld theory.

\medskip

Compact astrophysical objects, such as neutron stars, often harbour very strong magnetic fields. In fact, magnetars, a special subclass of neutron stars, have the strongest magnetic fields known in the universe \cite{TZW15,KB17}, estimated to reach up to $10^{11} \,\mathrm{T}$ at the star's surface. Such environments offer an opportunity for tests of nonlinear electromagnetic effects, complementary to those performed in particle colliders \cite{BR13,MCLP17,EBHRRM17,PCdL18} (a critical re-examination of ``quantum vacuum friction'' for the neutron star spin-down process has been recently presented in \cite{RH19}).

\medskip

The niche of NLE models has been heavily populated over the last three decades, based on various choices of NLE Lagrangian functions \cite{Soleng95,HM07,HM08,Hendi12,Hendi13,Kruglov15,Kruglov16,ISm18}. Large part of the motivation for this line of research comes from the quest for regular black hole solutions \cite{GSP84,deO94,ABG98,ABG99,ABG00,YT00,DARG09,RWX13}. Namely, just as the quantum corrections can regularize divergences related to classical point charges, it is expected that a resolution of black hole singularities \cite{AS11,AEP18,Juric18,GJSS19} might also appear within some of the NLE models (see, however, Bronnikov's constraints in \cite{Bronnikov00}). In order to understand consequences of nonlinearities of electromagnetic fields in general relativistic context better, it is important to analyse black hole exteriors immersed in such fields. Main objective of this paper is to find a NLE generalization of the Wald's solution on Schwarzschild spacetime, a magnetic field which is homogenoeus at spatial infinity and regular on the black hole horizon.

\medskip

The paper is organized as follows. In section 2 we overview complications introduced by nonlinearities in Maxwell's equations and why it is not straightforward to generalize Wald's solution for NLE. In section 3 we explain the details of perturbative approach to problem and in section 4 we present a solution on Schwarzschild black hole background. Section 5 is devoted to brief analysis of asymptotic properties of the field. In section 6 we present an alternative approach to the problem, via introduction of the magnetic scalar potential and in section 7 we discuss a related problem of (spherically symmetric, highly conducting) neutron star immersed in nonlinear magnetic field. In final section we briefly analyse the NLE generalization of the Wald's solution and highlight major open problems.

\medskip

\emph{Conventions and notation}. Basic electromagnetic invariants are defined as
\be\label{eq:FFGG}
\FF \equiv F_{ab} F^{ab} \qquad \textrm{and} \qquad \GG \equiv F_{ab}\,{*F}^{ab} \ .
\ee
One must be careful about the variations of the abbreviations used throughout the literature, where $\FF$ and $\GG$ might come with an extra factor, such as $\,\pm 1/4$. Derivatives of functions with respect to these variables are denoted by subscripts, such as $H_\FF = \dd_\FF H$, $H_\GG = \dd_\GG H$ and $H_{\FF\GG} = \dd_\GG \dd_\FF H$ for some function $H = H(\FF,\GG)$. We use subscript ``$\infty$'' for fields evaluated at infinity, that is in the limit when $r \to \infty$, while the subscript ``$0$'' is reserved for fields which are part of the basic Wald's solution. Unless stated otherwise, we use natural units with $c = G = 4\pi\epsilon_0 = \mu_0/4\pi = 1$.

\section{Nonlinear obstacles}

Large class of NLE models can be described by Lagrangian density $\LL(\FF,\GG)$, a sufficiently smooth function of electromagnetic invariants $\FF$ and $\GG$. Corresponding generalized Maxwell's equations can be written as 
\be
\df F = 0 \qqd \df\,{*Z} = 0
\ee
where $Z_{ab}$ is an auxiliary 2--form,
\be
Z = -4(\LL_\FF\,F + \LL_\GG \, {*F}) \ .
\ee
Can we still use Papapetrou's ansatz in this NLE context? The second generalized Maxwell's equation for $F = \df K$ is reduced to
\be
\df\LL_\FF \w {*F} - \df\LL_\GG \w F = 0 \ .
\ee
Furthermore, as
\be
\df\LL_\FF = \LL_{\FF\FF} \, \df\FF + \LL_{\FF\GG} \, \df\GG \quad \textrm{and} \quad \df\LL_\GG = \LL_{\GG\FF} \, \df\FF + \LL_{\GG\GG} \, \df\GG \ ,
\ee
here one has to deal with terms such as $\df\FF \w F$, $\df\GG \w F$, $\df\FF \w {*F}$ and $\df\GG \w {*F}$. For example, using an auxiliary vector field $X^a \equiv \nabla^a\FF$, we have
\be
*(\df\FF \w {*F}) = -i_X F = -i_X \df K = (\df i_X - \Lie_X) K \ .
\ee
Now, although
\be
i_X K = K^a \nab{a} \FF = \Lie_K \FF = 0 \ ,
\ee
and
\be
\Lie_X K^a = -\Lie_K X^a = -\Lie_K \nabla^a \FF = -g^{ab} \, \nab{a} \Lie_K\FF = 0 \ ,
\ee
nevertheless
\be
\Lie_X K_a = \Lie_X (g_{ab} K^b) = K^b \Lie_X g_{ab} \ ,
\ee
which in general doesn't have to vanish! There is even less hope that combination of all the terms mentioned above might ``conspire'' to cancel each other. The bottom line is that the basic ansatz $F = \df K$ simply does not work for general NLE and one has to find an alternative.

\medskip

One possible modification of the original idea is to use rescaled Killing vector field, so that $F = \df (\psi K)$ with some auxiliary function $\psi$. However, although we immediately have $\df F = 0$, the other equation $\df\,{*Z} = 0$ implies
\bea
(\LL_\FF \,{*\df K} + \df\LL_\GG \w K) \w \df\psi + \df\LL_\FF \w i_K {*\df\psi} + \nonumber\\
+ \LL_\FF ({* \df\Lie_K \psi} - (\Box\psi)\,{*K}) + (\df\LL_\FF \w {*\df K} - \df\LL_\GG \w \df K) \psi = 0 \ .
\eea
Main complication here comes from the fact that both invariants $\FF$ and $\GG$ are quadratic in $\psi$, thus for some general Lagrangian we are dealing with highly nonlinear differential equation for $\psi$. Unfortunately, we were not able to find a systematic approach for the exact solution of this problem. 

\medskip

In order to make some progress we resort to an approximation scheme, by looking at the perturbative expansion around original Wald's solution.

\section{Perturbative approach}

NLE Lagrangian densities considered throughout the literature are usually assumed to be a function that allows a double Taylor series expansion,
\be
\LL(\FF,\GG) = \sum_{m,n=0}^\infty c_{mn} \, \FF^m \GG^n
\ee
with some real coefficients $c_{mn}$. Without loss of generality one might assume here that $c_{00} = c_{01} = 0$, as these are non-dynamical terms. Also, for consistency with low-energy Maxwellian limit, we must take $c_{10} = -1/4$. CP--violating term $c_{11}$ \cite{MF09,BR13} has been recently constrained by the measurements at the ATLAS Collaboration \cite{AM18}. Here we are mainly interested in models with $c_{11} = 0$. Two most well-known examples are Euler--Heisenberg theory, with
\be\label{eq:EH}
\LL^{(\mathrm{EH})} = -\frac{1}{4}\,\FF + \frac{\alpha^2}{360 m_e^4} \left( 4\FF^2 + 7\GG^2 \right) + \dots
\ee
where $\alpha$ is the fine--structure constant and $m_e$ electron mass, and Born--Infeld theory,
\be\label{eq:BI}
\LL^{(\mathrm{BI})} = b^2 \left( 1 - \sqrt{1 + \frac{\FF}{2b^2} - \frac{\GG^2}{16b^4}} \right)
\ee
with parameter $b$ (effectively, the upper bound for electromagnetic field strength). Lagrangian density (\ref{eq:BI}) can be expanded as
\be
\LL^{(\mathrm{BI})} = -\frac{1}{4}\,\FF + \frac{1}{32b^2} \left( \FF^2 + \GG^2 \right) + \dots
\ee

\medskip

In what follows, we shall assume that electromagnetic Lagrangian density, expanded with respect to a physical coupling constant $\lambda$, has a form
\be
\LL(\FF,\GG) = -\frac{1}{4}\,\FF + \lambda \ell(\FF,\GG) + O(\lambda^2) \ .
\ee
For the gauge field 1--form $A_a$ we use the ansatz
\be
A_a = K_a + \lambda v_a + O(\lambda^2)
\ee
where $v_a$ is some unknown 1--form, perturbative correction to the basic Wald's solution. Consequently, electromagnetic 2--form is
\be\label{eq:F}
F = F_0 + \lambda \, \df v + O(\lambda^2) \ ,
\ee
with $F_0 \equiv \df K$. We already know that $\df F_0 = 0 = \df {*F}_0$, so that $\df F = 0$ is satisfied at the $O(\lambda^1)$ order. Let us look more closely at the second NLE Maxwell's equation, 
\be
\df (\LL_\FF \, {*F} - \LL_\GG \, F) = 0 \ .
\ee
Using expansions
\be
\LL_\FF = -\frac{1}{4} + \lambda \ell_\FF + O(\lambda^2) \qqd \LL_\GG = \lambda \ell_\GG + O(\lambda^2)
\ee
we get
\be
\df {*Z} = \lambda \left( \df {*\df} v - 4\,\df\ell_\FF \w {*\df} K + 4\,\df\ell_\GG \w \df K \right) + O(\lambda^2) \ .
\ee
Furthermore, using
\bea
\df\ell_\FF = & \ell_{\FF\FF}\,\df\FF + \ell_{\FF\GG}\,\df\GG \ , \\
\df\ell_\GG = & \ell_{\GG\FF}\,\df\FF + \ell_{\GG\GG}\,\df\GG
\eea
and expansions of the electromagnetic invariants,
\bea
\FF = & \FF_0 + 2\lambda (\df K)_{ab} (\df v)^{ab} + O(\lambda^2) \ , \\
\GG = & \GG_0 + 2\lambda ({*\df} K)_{ab} (\df v)^{ab} + O(\lambda^2) \ ,
\eea
we get the master equation for $v_a$,
\be\label{eq:master}
\df {*\df} v = {*J}_{\mathrm{eff}}
\ee
with ``effective 4-current'' $J_{\mathrm{eff}}^a$, such that
\be
{*J}_{\mathrm{eff}} = 4(\ell_{\FF\FF}\,\df\FF + \ell_{\FF\GG}\,\df\GG)_0 \w {*\df} K - 4(\ell_{\GG\FF}\,\df\FF + \ell_{\GG\GG}\,\df\GG)_0 \w \df K \ .
\ee
The ``$0$'' subscript above implies that terms in parenthesis have to be evaluated for the basic $F_0 = \df K$ ansatz. Just for consistency, it is straightforward to check that ${*J}_{\mathrm{eff}}$ is indeed a closed 3--form,
\be
\df {*J}_{\mathrm{eff}} = 0 \ .
\ee
Our main focus will be on the Euler--Heisenberg Lagrangian, with
\be
\ell^{(\mathrm{EH})} = 4\FF^2 + 7\GG^2 \qqd \lambda^{(\mathrm{EH})} = \frac{\alpha^2}{360 m_e^4} \ ,
\ee
and the Born-Infeld Lagrangian, with
\be
\ell^{(\mathrm{BI})} = \FF^2 + \GG^2 \qqd \lambda^{(\mathrm{BI})} = \frac{1}{32b^2} \ .
\ee
As in both of these examples we have $\ell_{\FF\GG} = 0$, that is $c_{11} = 0$, the master equation (\ref{eq:master}) reduces to
\be
\df {*\df} v = 4 (\ell_{\FF\FF}\,\df\FF)_0 \w {*\df} K - 4(\ell_{\GG\GG}\,\df\GG)_0 \w \df K \ .
\ee

Before we proceed, let us make several comments on the regime of applicability of the approximation scheme presented in this section. Namely, we assume that the electromagnetic field is strong enough to reveal nonlinear corrections, but still weak enough to allow test field approximation. In the absence of exact solutions, magnitude of order estimations in differential equations are often a slippery slope, but still one can hope that basic relevant information might be extracted if all relevant scales are taken into account. For example, if the Einstein's tensor $G_{ab}$ is of order $L_g^{-2}$, where $L_g$ is a characteristic gravitational length scale for the problem, while the energy density of the magnetic field is $\mathbf{B}^2/2\mu_0$, then the weak field condition, based on comparison of the left and right hand sides of the Einstein gravitational field equation, might be written as $L_g^{-2} \gg 4\pi G \mathbf{B}^2/c^4\mu_0$. A sensible choice for $L_g$ would be Schwarzschild radius, $L_g \sim 3(M/M_\odot) \cdot 10^3\,$m, where $M_\odot$ is the Solar mass. This gives us condition $|\mathbf{B}| \ll (M_\odot/M) \cdot 10^{15}\,$T, which indicates that even the strongest known magnetic fields can be treated as test fields, as long as the black hole mass $M$ is below the order of $10^4\,M_\odot$ (similar discussion can be found in \cite{Petri15b}).

\section{Setting the problem upon the Schwarzschild spacetime}

Schwarzschild spacetime metric is a static, spherically symmetric solution of vacuum Einstein equation \cite{Wald},
\be\label{eq:Sch}
\df s^2 = -f(r) \, \df t^2 + \frac{\df r^2}{f(r)} + r^2 \left( \df\theta^2 + \sin^2\theta \, \df\varphi^2 \right)
\ee
with
\be
f(r) = 1 - \frac{2M}{r} \ .
\ee
Schwarzschild spacetime possesses two Killing vector fields, stationary $k = \dd/\dd t$ and axial $m = \dd/\dd\varphi$. In general we might start with the Killing vector field
\be
K^a = \alpha k^a + \beta m^a \ ,
\ee
with some real constants $\alpha$ and $\beta$. Corresponding electromagnetic invariants, evaluated for $F_0 = \df K$, are given by
\be
\FF_0 = -\frac{8M^2}{r^4} \, \alpha^2 + 8 \left( 1 - \frac{2M}{r} \, \sin^2\theta \right) \beta^2
\ee
and
\be
\GG_0 = -16M\,\frac{\cos\theta}{r^2}\,\alpha\beta \ .
\ee
As in the Wald's solution for the Schwarzschild case, we shall focus on the choice $\alpha = 0$, which will \emph{a posteriori} prove to be appropriate for our problem (note that in the Wald's solution parameter $\alpha$ is proportional to the angular momentum $J$). Here we have an important simplification $\GG_0 = 0$, so that equation (\ref{eq:master}) reduces even further to
\be\label{eq:dvSch}
\df {*\df} v = 4\beta(\ell_{\FF\FF}\,\df\FF)_0 \w {*\df} m \ .
\ee
Direct calculation gives
\be
\df\FF_0 \w {*\df} m = \frac{32M\beta^2 \sin\theta}{r} \, \left( f(r)\sin^2\theta - 2\cos^2\theta \right) \df t \w \df r \w \df\theta \ .
\ee
Symmetries of the problem suggest that an appropriate ansatz is of the form $v = h(r,\theta)\,\df\varphi$. This allows us to find a solution of the equation (\ref{eq:dvSch})
\be\label{eq:vC}
v = C \Big( 4(2r - 5M) \cos(2\theta) + (M-2r) \big( 3 + \cos(4\theta) \big) \Big) \, \df\varphi \ ,
\ee
with a constant $C$. As
\be
\df {*\df v} = \frac{64C\sin\theta}{r} \, \left( f(r)\sin^2\theta - 2\cos^2\theta \right) \df t \w \df r \w \df\theta
\ee
it follows that $C = 2\beta^3 M (\ell_{\FF\FF})_0$. The remaining constant $\beta$ can be fixed from boundary conditions, as discussed below.

\section{Asymptotia}

We want to make sure that the perturbative solution found in the previous section is such that (a) asymptotically represents homogeneous magnetic field, and (b) corresponding electric and magnetic Komar charges remain zero at the $O(\lambda^1)$ level.

\medskip

Homogeneous magnetic field in Minkowski spacetime can be written as $B_\infty\,\df z = B_\infty\,\df(r\cos\theta)$, with constant $B_\infty$, and the corresponding electromagnetic 2--form is
\be\label{eq:homB}
F_\infty = B_\infty \big( r\sin^2\theta \, \df r \w \df\varphi + r^2\cos\theta\sin\theta \, \df\theta \w \df\varphi \big) \ .
\ee
Wald's solution in Schwarzschild is given by 2--form
\be\label{eq:SchWald}
F_0 = \frac{1}{2}\,B_\infty\,\df m = B_\infty \big( r\sin^2\theta \, \df r \w \df\varphi + r^2\cos\theta\sin\theta \, \df\theta \w \df\varphi \big) \ .
\ee
Formally, this has exactly the same functional form as (\ref{eq:homB}). As Schwarzschild spacetime metric is asymptotically flat, this immediately proves that field (\ref{eq:SchWald}) asymptotically represents homogeneous magnetic field.

\medskip

In the NLE case one has to check behaviour of the 1--form $v_a$ at spatial infinity. As
\bea
\df v = & -32\beta^3 M (\ell_{\FF\FF})_0 \, \big( \sin^4\theta \, \df r \w \df\varphi + \nonumber \\
 & + (2r - 5M + (M-2r)\cos(2\theta))\sin\theta\cos\theta \, \df\theta \w \df\varphi \big)
\eea
we have
\be
\lim_{r\to\infty} \frac{(\df v)_{r\varphi}}{(F_0)_{r\varphi}} = 0 \ , \ \ \textrm{and} \quad \lim_{r\to\infty} \frac{(\df v)_{\theta\varphi}}{(F_0)_{\theta\varphi}} = 0 \ ,
\ee
so that 2--form $F = F_0 + \lambda\df v$ asymptotically behaves as Wald's $F_0$. Also, note that corresponding vector field $v^a$, unlike gauge field $A^a = K^a$, vanishes at infinity, $\lim_{r\to\infty} v^{\mu} = 0$.

\medskip

All this allows us to choose normalization just as in the Wald's solution, $\beta = B_\infty/2$, so that finally
\be\label{eq:v}
v = \frac{(\ell_{\FF\FF})_0}{4}\, B_\infty^3 M \Big( 4(2r - 5M) \cos(2\theta) + (M-2r) \big( 3 + \cos(4\theta) \big) \Big) \, \df\varphi \ .
\ee

\medskip

Electric charge $Q_\mathcal{S}$ and magnetic charge $P_\mathcal{S}$ enclosed by a smooth closed 2-surface $\mathcal{S}$ are given by Komar integrals,
\be
Q_\mathcal{S} = \frac{1}{4\pi} \, \oint_\mathcal{S} {*Z} \ , \quad \mathrm{and} \quad P_\mathcal{S} = \frac{1}{4\pi} \, \oint_\mathcal{S} F \ .
\ee
Now, we know that both $Q_\infty$ and $P_\infty$ for Wald's solution are zero by construction. Using the expansion
\be
{*Z} = {*F}_0 + \Big( 4(-\ell_\FF {*F} + \ell_\GG F)_0 + {*\df v} \Big) \lambda + O(\lambda^2)
\ee
and the fact that $\ell_\FF = 2\FF$, $\ell_\GG = 2\GG$, $\lim_{r\to\infty} \FF_0 = 8\beta^2$, $({*F_0})_{\theta\varphi} = 0$ and $({*\df v})_{\theta\varphi} = 0$, electric charge $Q_\infty$ remains unaltered in our solution at the $O(\lambda^1)$ order. Furthermore, $(\df v)_{\theta\varphi}$ contains $\sin(2\theta)$ and $\sin(4\theta)$ parts, both of which vanish upon integration over the interval $[0,\pi]$, so that magnetic charge $P_\infty$ also remains unaltered, that is zero.

\section{Scalar potentials}

Just as in classical electrostatics and magnetostatics, a useful strategy for problem solving is introduction of scalar potentials, whenever this is possible \cite{ISm12,ISm14,GS17}. Magnetic field 1--form defined with respect to a vector field $X^a$ is given by $B[X]_a \equiv X^b\,{*F}_{ba}$. A convenient choice for $X^a$ is a Killing vector field $K^a$: Assuming that electromagnetic field is  symmetry inheriting \cite{Woo73a,Woo73b,MW75,Coll75,RT75,WY76a,WY76b,Tod06,CDPS16,BGS17}, $\Lie_K F_{ab} = 0$, solution of source-free Maxwell's equations, corresponding magnetic 1--form will be closed,
\be
\df B[K] = \df i_K {*F} = (\Lie_K - i_K \df) \, {*F} = 0 \ .
\ee
Furthermore, if the black hole exterior is simply connected, then there is a globally defined function $\Psi$, magnetic scalar potential, such that $B[K] = -\df\Psi$. Just as the surface gravity, the potential $\Psi$ is also constant over a Killing horizon \cite{ISm12,ISm14}. For example, magnetic field for Wald's solution, defined with respect to the Killing vector field $k = \dd/\dd t$, is
\be
B_0[k] = B_\infty \left( \cos\theta \, \df r - r f(r) \sin\theta \, \df\theta \right) \ ,
\ee
and, up to constant, corresponding scalar potential is
\be\label{eq:WaldPsi}
\Psi_0 = -B_\infty f(r) \, r\cos\theta \ .
\ee
The gauge choice implicitly used here is such that potential vanishes at the horizon, $\lim_{r\to 2M} \Psi_0 = 0$. At the spatial infinity we have $\lim_{r\to\infty} \Psi_0 = -B_\infty z$.

\medskip

In NLE the magnetic field 1--form $B[K]_a$ is no longer necessarily closed, but one might introduce another, ``nonlinear $H$--field'' $H[K]_a \equiv K^b\,{*Z}_{ba}$, which is closed \cite{BGS17} by analogous reasoning,
\be
\df H[K] = \df i_K {*Z} = (\Lie_K - i_K \df) \, {*Z} = 0 \ .
\ee
This allows us to introduce NLE magnetic scalar potential $\Upsilon$, via $H[K] = -\df\Upsilon$. Constancy of the potential $\Upsilon$ over a Killing horizon was recently discussed in \cite{BGS17}.

\medskip

On static spacetime Maxwell's equations imply a divergence equation \cite{Heusler}
\be\label{eq:divB}
\nab{a} \left( \frac{B[k]^a}{N} \right) = 0 \ ,
\ee
where $N \equiv k^a k_a$. From here we immediately have the equation for the scalar potential. For example, for axially symmetric potential $\Psi$ in Schwarzschild spacetime it reads
\be\label{eq:L}
L[\Psi] \equiv f(r)\,\frac{\dd}{\dd r} \left( r^2 \, \frac{\dd\Psi}{\dd r} \right) + \frac{1}{\sin\theta} \, \frac{\dd}{\dd\theta} \left( \sin\theta \, \frac{\dd\Psi}{\dd\theta} \right) = 0 \ ,
\ee
where we have, in order to simplify some equations below, introduced a differential operator $L$. Note that $L[\Psi] = -r^2 f(r) \nabla^a ((\nab{a}\Psi)/N)$. Partial differential equation (\ref{eq:L}) allows a separation of variables via $\Psi(r,\theta) = R(r)P(\cos\theta)$. The $\theta$--part comes out, not surprisingly, as a solution of the Legendre differential equation, while the radial part is a function of the form
\be
R_\ell(r) = \left( \frac{r}{2M} - 1 \right) \left( a_\ell P'_\ell\left(\frac{r}{M} - 1\right) + b_\ell Q'_\ell\left(\frac{r}{M} - 1\right) \right) \ ,
\ee
with Legendre polynomial $P_\ell$ and Legendre function of the second kind $Q_\ell$. Some of the earliest treatments of these solutions can be traced back to 1960s \cite{GO65,Israel67} and early 1970s \cite{AC70,CW71}.

\medskip

The NLE case is considerably more delicate and, in order to simplify matters, we shall reach for some additional assumptions. Despite the fact that (\ref{eq:divB}) still holds in the NLE case, as the magnetic scalar $\Upsilon$ is defined with respect to the $H$--field, we need to find the corresponding divergence equations. Our focus will be on solution which are ``purely magnetic'' in a sense that $k^b F_{ab} = 0$. In this case we have a useful relation
\be\label{eq:BH}
H[k]_a = -4k^b\,{*Z}_{ba} = -4\LL_\FF\,k^b\,{*F}_{ba} = -4\LL_\FF B[k]_a
\ee
which can be used in (\ref{eq:divB}) to get
\be\label{eq:divH}
\nab{a} \left( \frac{H[k]^a}{N\LL_\FF} \right) = 0 \ .
\ee
Again, as above, we resort to perturbative approach, by expanding everything with respect to coupling constant $\lambda$. In order to simplify notation, we assume that $\ell = p\FF^2 + q\GG^2$, with some real constants $p$ and $q$. Note that $(p,q) = (4,7)$ in the Euler--Heisenberg case, $(p,q) = (1,1)$ in the Born--Infeld case, and $\ell_{\FF\FF} = 2p$ in both of them. First of all we have
\be
\Upsilon = \Psi_0 + \lambda \Psi_1 + O(\lambda^2) \ ,
\ee
and, after some algebra,
\be
\nab{a} \left( \frac{H^a}{N} \right) - 16p\lambda \nab{a} \left( \frac{H_b H^b}{N^2}\,H^a \right) + O(\lambda^2) = 0 \ ,
\ee
where, for simplicity, we have suppressed the argument ``$[k]$''. This gives us back the zeroth order equation $\nabla^a ((\nab{a}\Psi_0)/N) = 0$ and the equation for the perturbation,
\be\label{eq:pertPsi}
\nabla^a \left( \frac{\nab{a}\Psi_1}{N} \right) = 16p \, \nabla^a \left( \frac{(\nab{b}\Psi_0)(\nabla^b\Psi_0)}{N^2}\,\nab{a}\Psi_0 \right) \ .
\ee
More concretely, if we insert Wald's solution (\ref{eq:WaldPsi}), equation (\ref{eq:pertPsi}) becomes
\be
L[\Psi_1] = 48pM B_\infty^3 f(r) \sin(2\theta) \sin\theta \ .
\ee
For this problem one may use an ansatz of the form 
\be
\Psi_1(r,\theta) = f(r) \Big( a(r) + b(r)\cos(2\theta) \Big) \cos\theta
\ee
and, by choice of integration constants, discard part of the solution that grows faster than $O(r^1)$ at spatial infinity. Finally, this gives us
\be\label{eq:WaldPsi1}
\Psi_1(r,\theta) = 4p B_\infty^3 \, f(r) \Big( 4r - 5M + M\cos(2\theta) \Big) \cos\theta \ .
\ee
It is straightforward but tedious exercise to check that the electromagnetic field given by this scalar potential  indeed agrees with the previously obtained correction (\ref{eq:v}) to Wald's solution.

\section{Remarks on neutron stars}

The analysis above assumes that a black hole is present in spacetime, and part of the boundary conditions is regularity of the fields at the black hole horizon. Somewhat different situation appears if instead of a black hole we have a star. Here we are looking at an idealized model of a relativistic, spherically symmetric and highly conducting star. Although the electric conductivity in different parts of a neutron star may significantly vary, from non-superconducting outer layers to superconducting core \cite{CH08,Potekhin10,Glendenning}, we shall simply assume that the whole star is represented by a ball of infinite electric conductance. On top of all this, we shall initially strip the star of its internal magnetic field (which may be subsequently superposed for slightly more realistic model) and immerse it in external test homogeneous magnetic field, just as we did with the black hole.

\medskip

Superconducting materials in laboratory exhibit the Meissner effect, expulsion of external magnetic field. If a superconducting ball of radius $R$ is placed in a homogeneous magnetic field of strength $B_\infty$, the resulting field is a superposition of the external field and a dipole magnetic field produced by induced surface currents. Induced magnetic dipole can be found from classical junction condition, continuity of normal component of magnetic field at boundary surface, and in flat, Minkowski case is given by $\mu = -B_\infty R^3/2$.

\medskip

Let us now turn to a more general case of a static spacetime. We assume that spacetime can be foliated by diffeomorphic ``equal time'' hypersurfaces $\Sigma$, each of which contains a compact spacelike 2-surface $\mathcal{S} \subseteq \Sigma$, such as a boundary of a star, with normal $n^a$. The divergence relation (\ref{eq:divB}) allows us to deduce a junction condition for the magnetic field at $\mathcal{S}$. Assuming that square of the Killing vector $N$ is continuous at $\mathcal{S}$, it follows that the normal component of magnetic field, $n^a B_a$, has to be continuous at $\mathcal{S}$ as well. If, in addition, magnetic field vanishes in part of the spacetime bounded by $\mathcal{S}$ (star's interior) then we know that in fact $n^a B_a = 0$ at $\mathcal{S}$, and the scalar potential $\Psi$ satisfies Neumann boundary condition, $n^a \nab{a} \Psi = 0$ at $\mathcal{S}$.

\medskip

Back in the 1960s, in a precursor to no-hair theorems, Ginzburg and Ozernoy \cite{GO65} have analysed the magnetic dipole field in Schwarzschild spacetime. Part of the solutions, discussed in section 6, corresponding to this field is given by the $\ell=1$ term in the scalar potential,
\be
\Psi_{\mathrm{GO}}(r,\theta) = \frac{3\mu}{(2M)^2} \left( 1 + f(r) + \frac{r}{M}\,f(r)\ln f(r) \right)\cos\theta \ .
\ee
Expansion for large $r$ reveals classical potential on Minkowski background in the lowest order term, 
\be
\Psi_{\mathrm{GO}}(r,\theta) = \Big( r^{-2} + O(r^{-3}) \Big) \mu \cos\theta \ .
\ee
As we seek for the asymptotically homogeneous field, we may simply add Wald's solution,
\be
\Psi = \Psi_0 + \Psi_{\mathrm{GO}} \ .
\ee
Neumann boundary condition has to be imposed on the surface of our superconducting ball of radius $R > 2M$,
\be
\frac{\dd\Psi(R,\theta)}{\dd r} = 0 \ ,
\ee
from where one may find the induced magnetic dipole moment, 
\be
\mu = \frac{(2M)^2 B_\infty}{3} \left( \frac{3 - f(R)}{R} + \frac{1}{M}\,\ln f(R) \right)^{\!-1} \ .
\ee
We are not aware if this result was discussed previously in the literature. If one looks at the dipole moment $\mu$ as a function of mass $M$, its Taylor series around $M = 0$ reads
\be
\mu(M) = -\frac{B_\infty R^3}{2} + \frac{3B_\infty R^2}{4}\,M + O(M^2) \ ,
\ee
in agreement with the flat case, as $\lim_{M\to 0} \mu(M) = -B_\infty R^3/2$. Furthermore, as Maxwell's equations are linear, we might bring back the internal star's magnetic field simply by superposing it with the solution obtained here. For example, if the star's magnetic field is modelled by the dipole field, effectively we just have to replace the magnetic dipole moment $\mu$ with some novel $\widetilde{\mu}$. 

\medskip

Now, one might ask what happens if we take into account nonlinear electromagnetic effects? First we have to carefully re-examine junction conditions. In ``purely magnetic'' case 1--forms $B[k]_a$ and $H[k]_a$ are related by equation (\ref{eq:BH}). Assuming that $\LL_\FF$ is finite at $\mathcal{S}$, vanishing of $n^a B_a = 0$ at the superconducting boundary $\mathcal{S}$ implies that $n^a H_a = 0$ and $n^a \nab{a} \Upsilon = 0$ at $\mathcal{S}$. 

\medskip

If we write the basic solution as $\Psi = R(r)\cos\theta$, the linearized equation for the potential (\ref{eq:pertPsi}) takes the form
\be
L[\Psi_1] = -\frac{4p\sin(2\theta)}{f(r)}\,(\rho_+(r) + \rho_-(r)\cos(2\theta)) \ ,
\ee
where we have introduced two auxiliary functions,
\bea
\rho_\pm(r) = & \pm \Big( r(r-2M)R'' - 2MR' - 4R \Big)R^2 + \nonumber \\
 & + r(r-2M) \Big( (r-2M)(3rR'' + 2R') + (-2 \pm 4) R \Big)R'^2 \ .
\eea
This is a linear, nonhomogeneous partial differential equation, with known homogeneous solutions (see section 6). Usual technique for the particular solution includes integration of the associated Green's function (see \cite{CW71,Persides73,FL17}) with the inhomogeneity. However, in this case the result is an infinite series, where each term (evaluated with help of the package \emph{Mathematica}) is itself a nontrivial sum of over a hundred of functions, combination of polynomials, logarithms and polylogarithms in radial coordinate. Written in this way, solution becomes completely intractable and it is highly nontrivial to impose boundary conditions. It remains an open question if this problem can be solved in a closed, analytical form.

\medskip

We note in passing that the analysis in \cite{Petri15a,Petri15b} is somewhat related as it treats the QED corrections (modelled by the Euler--Heisenberg Lagrangian) of magnetic dipole on spherically symmetric neutron star, albeit with completely different formalism.

\section{Discussion}

\begin{figure}[htb]
\centering
\includegraphics[scale=.6]{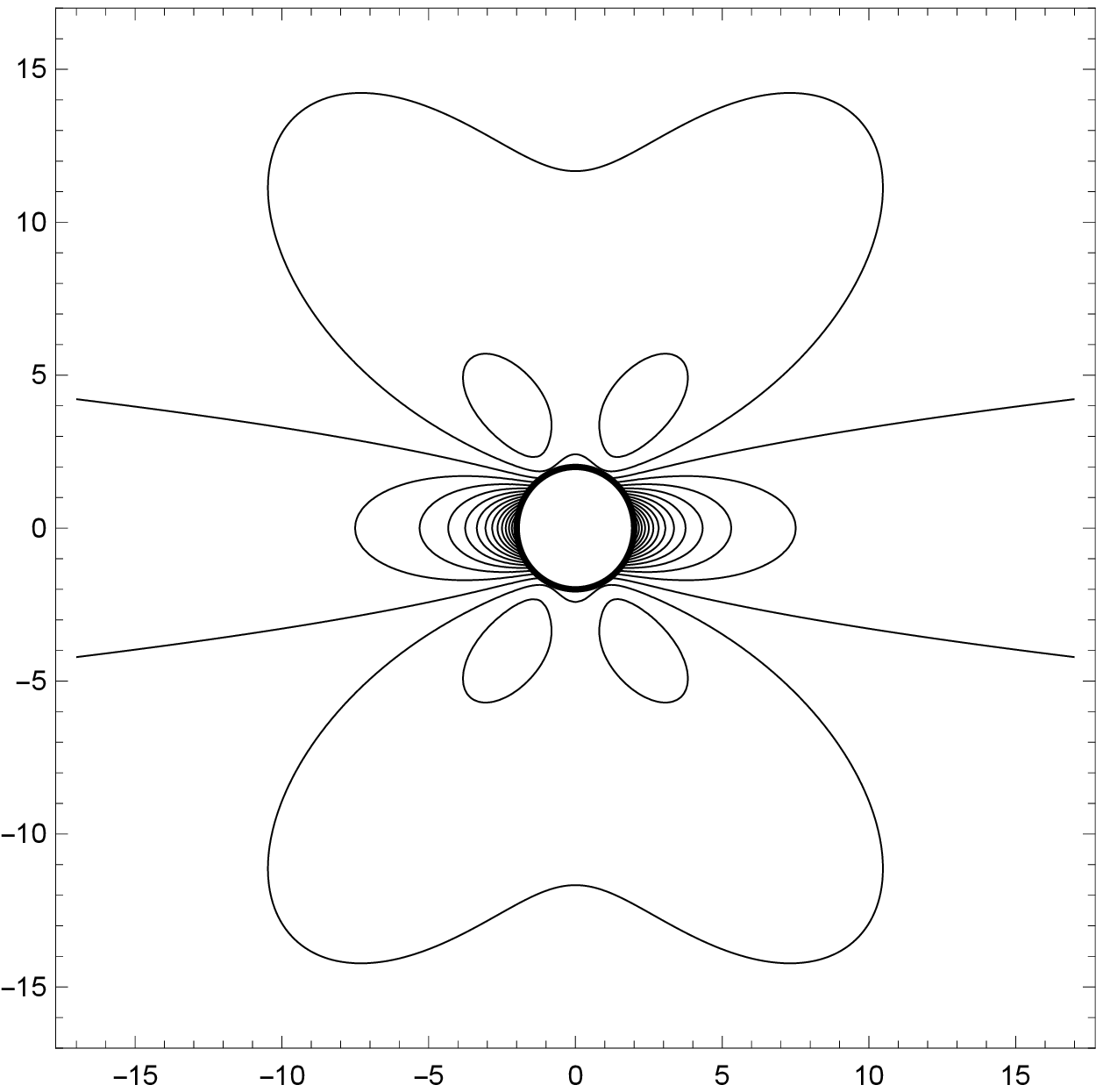}
\includegraphics[scale=.6]{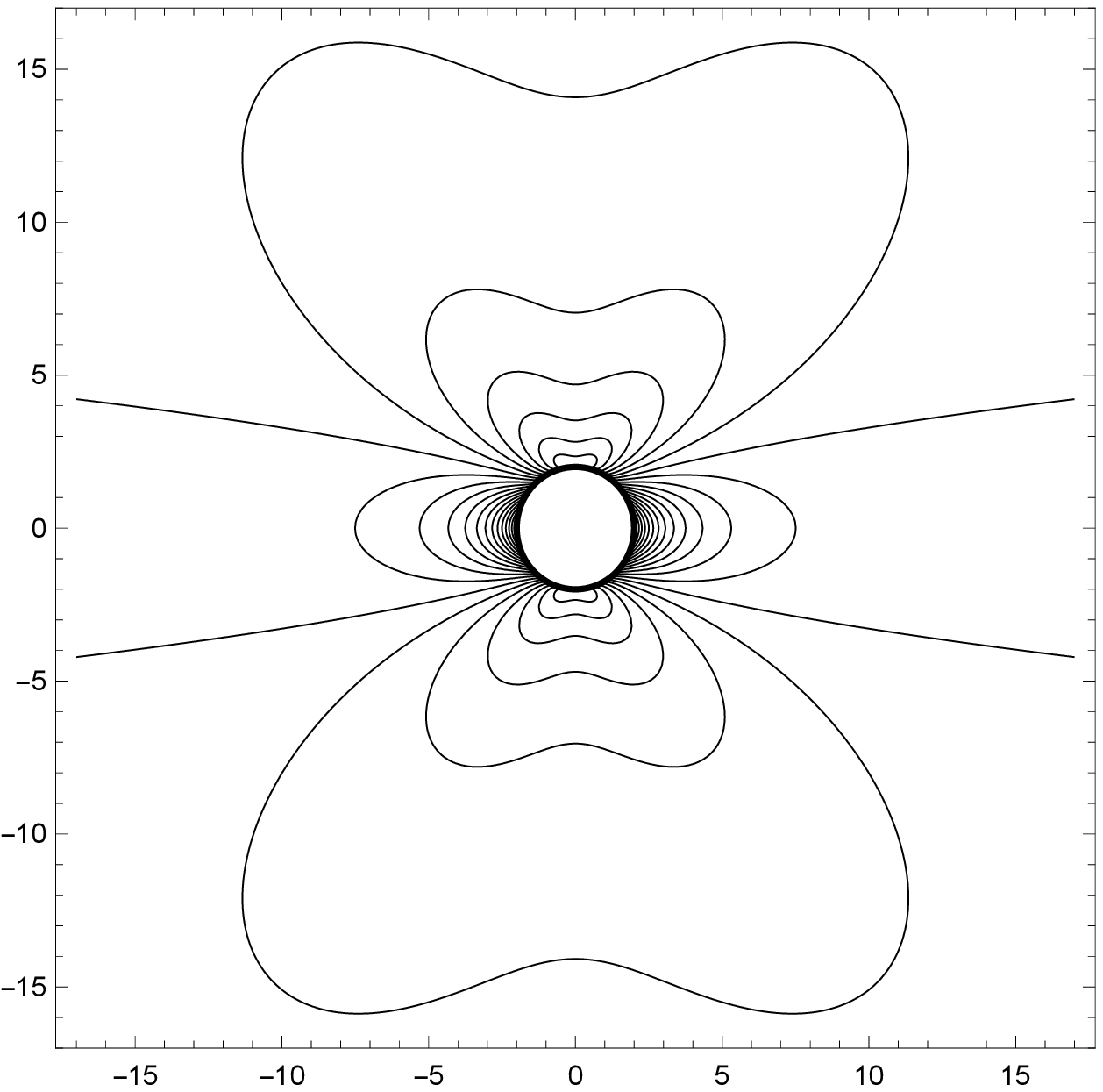}

\caption{Contour plots in $r$--$\theta$ plane for $M=1$ (black hole horizon is denoted by the black circle in the middle). Left: Contour plot of correction $\widehat{\FF}_1$. Right: Contour plot of rescaled relative correction $8\beta^2\,\widehat{\FF}_1/\FF_0$.}
\label{fig:graph}
\end{figure}

Correction to Wald's solution may be represented in multitude ways. If one expands magnetic field, defined with respect to the Killing vector field $k^a$,
\be
B[k] = B_0[k] + \lambda \, B_1[k] \qqd B_1[k] \equiv i_k {*\df v} \ ,
\ee
we have explicitly
\bea
B_1[k] = & 4M(\ell_{\FF\FF})_0 B_\infty^3 \Big( f(r) \sin^3\theta \, \df\theta - \nonumber\\
 & - \frac{\cos\theta}{r} \left( 2r - 5M + (M-2r)\cos(2\theta) \right) \df r \Big)
\eea
Physical magnetic field, on the other hand, is the one measured by some concrete physical observer (measuring apparatus). For example, for the static observer with 4-velocity $u^a = k^a/\sqrt{-N}$ we have
\be
B[u]^a = \frac{1}{\sqrt{f(r)}}\,B[k]^a \ .
\ee
Still, we find that it is better to look first at the observer independent quantities, such as electromagnetic invariants. Correction to the first electromagnetic invariant may be decomposed as $\FF = \FF_0 + \delta\FF$. Just to put all the prefactors aside we introduce
\be
\widehat{\FF}_1 \equiv -\frac{1}{16B_\infty^3 M(\ell_{\FF\FF})_0} \, (\df m)_{ab} (\df v)^{ab} \ ,
\ee
so that
\be
\delta\FF = -16\lambda B_\infty^4 M(\ell_{\FF\FF})_0 \, \widehat{\FF}_1 + O(\lambda^2) \ .
\ee
Direct calculation gives
\be
\widehat{\FF}_1 = \frac{1}{r}\,f(r) \sin^4\theta + \frac{\cos^2\theta}{r} \left( (3+f(r))\sin^2\theta - 2(1-f(r)) \right) \ .
\ee
The solution is regular on the black hole horizon, as $\widehat{\FF}_1$ remains bounded for $r \to 2M$. Contour plots for $\widehat{\FF}_1$ can be seen on Figure \ref{fig:graph}.

\bigskip

\bigskip

The picture reveals an interesting feature, local maxima of $\widehat{\FF}_1$ along two circles at $(r_c,\theta_\pm)$. Analytically, from $\dd_r \widehat{\FF}_1 = 0$ and $\dd_\theta \widehat{\FF}_1 = 0$, one gets respectfully
\bea
48M - 7r + 4(r + 4M)\cos(2\theta) + 3r\cos(4\theta) = 0 \ ,\\
(r + 2M + 3r\cos(2\theta))\sin(2\theta) = 0 \ .
\eea
This system of equations may be simplified with a substitution $x = \cos(2\theta)$, leading to a solution (here we are looking only at the black hole exterior, $r > 2M$)
\be
r_c = \frac{4 + \sqrt{13}}{2}\,M \qqd \cos(2\theta_\pm) = \frac{4\sqrt{13} - 19}{9} \ .
\ee
Approximately, these are $r_c \approx 3.8 M$, $\theta_+ \approx 60.3^\circ$ and, as $\cos(2(\pi-\theta)) = \cos(2\theta)$, $\theta_- \approx 119.7^\circ$. It would be interesting to see if this local maxima has some ramifications on trajectories of charged particles around the black hole, opening an opportunity for astrophysical tests.  

\medskip

A recent pair of papers \cite{TSSA18,TSA18} treat electromagnetic perturbations of static, spherically symmetric, charged black holes, bearing some resemblance to the analysis presented here. However, these papers are mainly focused on study of quasinormal modes (thus, different asymptotic boundary conditions) with less general class of NLE Lagrangians, $\LL = \LL(\FF)$.

\medskip

What happens if we have a NLE model with a $c_{11}$ term? Assuming that we still have $\GG_0 = 0$, the additional term on the right hand side of the master equation (\ref{eq:master}) is proportional to
\be
\df\FF_0 \w \df m = 96\beta^2 M \sin^3\theta \cos\theta \, \df r \w \df\theta \w \df\varphi \ .
\ee
Using a generalized ansatz, $A_a = \beta m_a + \lambda (v_a + \widetilde{v}_a)$, we were able to find a solution
\be
\widetilde{v} = 2M f(r) (\cos(3\theta) - 9\cos\theta) \, \df t \ .
\ee
Although this correction does not affect asymptotic homogeneous magnetic field, nor does it alter the zero values of charges $Q_\infty$ and $P_\infty$, it however introduces the electric field in a sense that in general $k^b F_{ab} \ne 0$ throughout the spacetime, even as $r \to \infty$. A natural step forward is to look at the further generalization with $K^a = \alpha k^a + \beta m^a$, but as this introduces $\GG_0 \ne 0$, equations become considerably more complicated and we leave this line of research for the future work. 

\medskip

Finally, two most important open questions that remain are (1) NLE perturbations of the neutron star immersed in homogeneous magnetic field (partially solved in section 7 above), and (2) generalization of this whole analysis for rotating compact objects, first and foremost Kerr black hole immersed in NLE environment. The rotating case is, as usual, a formidable task, which in the case of NLE perturbations can be easily demonstrated by complexity of invariants $\FF_0$ and $\GG_0$ evaluated on the Kerr spacetime, which directly translates into complexity of the ``effective 4-current'' $J^a$ in the master equation (\ref{eq:master}). Investigation of the possible alternative techniques (see e.g.~\cite{BBGD86}) of generalization of these solutions to non-static cases is left for the future work.

\ack
We would like to thank Tajron Juri\'c for careful reading of the manuscript and series of useful remarks.

\appendix
\section{Some identities from differential geometry}

Suppose that $(M,g_{ab})$ is a smooth Lorentzian manifold. The Hodge dual of a $p$--form $\omega_{a_1 \dots a_p}$ is defined as
\be
(*\omega)_{a_{p+1} \dots a_m} = \frac{1}{p!}\,\omega_{a_1 \dots a_p} \tensor{\epsilon}{^{a_1}^{\dots}^{a_p}_{a_{p+1}}_{\dots}_{a_m}} \ ,
\ee
while twice applied Hodge dual results in 
\be
{{**}\,\omega} = (-1)^{p(m-p)+1} \, \omega \ .
\ee
Contraction of a $p$--form $\omega_{a_1 \dots a_p}$ with a vector $X^a$ is defined by
\be
(i_X \omega)_{a_1 \dots a_{p-1}} = X^b \, \omega_{b a_1 \dots a_{p-1}} \ .
\ee
Calculations can often be simplified by ``flipping over the Hodge'',
\be
i_X {*\alpha} = {*(\alpha \w X)} \ ,
\ee
with a slight abuse of notation: the $X$ on the right hand side denotes the 1--form $X_a = g_{ab} X^b$ associated with the vector $X^a$. For a smooth vector field $X^a$ we have the Cartan's identity
\be
\Lie_X \, \omega = (i_X \df + \df i_X)\omega \ .
\ee
The Lie derivative commutes with the exterior derivative, $\Lie_X \df\omega = \df\Lie_X \omega$, while the Lie derivative with respect to a Killing vector field $K^a$ commutes with the Hodge dual, $\Lie_K\,{*\omega} = {*\Lie_K \omega}$.

\section{Useful bits}

Throughout the calculations one has to repeatedly use some Hodge duals, so it is useful to collect them in one place,
\be
{*(\df t \w \df r)} = -r^2 \sin\theta \, \df\theta \w \df\varphi \qqd {*(\df\theta \w \df\varphi)} = \frac{1}{r^2 \sin\theta} \, \df t \w \df r
\ee
\be
{*(\df t \w \df\theta)} = \frac{\sin\theta}{f(r)} \, \df r \w \df\varphi \qqd {*(\df r \w \df\varphi)} = -\frac{f(r)}{\sin\theta} \, \df t \w \df\theta
\ee
\be
{*(\df t \w \df\varphi)} = -\frac{1}{f(r)\sin\theta} \, \df r \w \df\theta \qqd {*(\df r \w \df\theta)} = f(r)\sin\theta \, \df t \w \df\varphi
\ee
Also, we have
\be
\frac{1}{2}\,\df m = r\sin^2\theta \, \df r \w \df\varphi + r^2\cos\theta\sin\theta \, \df\theta \w \df\varphi \ ,
\ee
and
\be
\frac{1}{2}\,{*\df m} = \cos\theta \, \df t \w \df r - r f(r) \sin\theta \, \df t \w \df\theta \ .
\ee
For 1--form $w = v/C$, where $v_a$ is the solution given by (\ref{eq:vC}), we have
\bea
\df w = & -16\sin^4\theta \, \df r \w \df\varphi - \nonumber\\
 & - 8 \sin(2\theta) \Big( 2r - 5M + (M-2r)\cos(2\theta) \Big) \df\theta \w \df\varphi \ ,
\eea
and
\bea
{*\df w} = & 16f(r)\sin^3\theta \, \df t \w \df\theta - \nonumber\\
 & - 16\,\frac{\cos\theta}{r^2} \Big( 2r - 5M + (M-2r)\cos(2\theta) \Big) \df t \w \df r \ .
\eea

\bigskip

\section*{References}


\bibliographystyle{iopnum}
\bibliography{nlepapw}

\end{document}